\documentclass[10pt,conference,letterpaper]{IEEEtran}
\usepackage{citesort}
\usepackage{color}
\usepackage[latin9]{inputenc}
\usepackage{amsthm}
\usepackage{amsmath}
\usepackage{epstopdf}
\usepackage{amsthm}
\usepackage{amssymb}
\usepackage{graphicx}
\usepackage{algorithmic}
\usepackage{algorithm}
\usepackage{mathrsfs}

\newtheorem{theorem}{Theorem}
\newtheorem{lemma}{Lemma}
\newtheorem{proposition}[theorem]{Proposition}
\theoremstyle{definition}
\newtheorem{definition}{Definition}

\theoremstyle{remark}
\newtheorem{remark}{Remark}

\newcommand{\PU}{\ensuremath {\mathit{PU}}}
\newcommand{\n}{\ensuremath {\mathit{n}}}
\newcommand{\SU}{\ensuremath {\mathit{SU}}}

\begin{document}

\title{Exploiting Wideband Spectrum Occupancy Heterogeneity for Weighted Compressive Spectrum Sensing}

\author{Bassem Khalfi$^{*}$, Bechir Hamdaoui$^{*}$, Mohsen Guizani$^{\dag}$, and Nizar Zorba$^{\dag}$\\
$^{*}$Oregon State University, $^{\dag}$Qatar University\\
$^{*}$\{khalfib,hamdaoui\}@eecs.orst.edu, $^{\dag}$mguizani@ieee.org, $^{\dag}$nizarz@qu.edu.qa}
\maketitle
\begin{abstract}
Compressive sampling has shown great potential for making wideband spectrum sensing possible at sub-Nyquist sampling rates. As a result, there have recently been research efforts that aimed to develop techniques that leverage compressive sampling to enable compressed wideband spectrum sensing. These techniques consider homogeneous wideband spectrum, where all bands are assumed to have similar PU traffic characteristics. In practice, however, wideband spectrum is not homogeneous, in that different spectrum bands could have different PU occupancy patterns. In fact, the nature of spectrum assignment, in which applications of similar types are often assigned bands within the same block, dictates that wideband spectrum is indeed heterogeneous, as different application types exhibit different behaviors. In this paper, we consider heterogeneous wideband spectrum, where we exploit this inherent, block-like structure of wideband spectrum to design efficient compressive spectrum sensing techniques that are well suited for heterogeneous wideband spectrum. We propose a weighted $\ell_1-$minimization sensing information recovery algorithm that achieves more stable recovery than that achieved by existing approaches while accounting for the variations of spectrum occupancy across both the time and frequency dimensions. Through intensive numerical simulations, we show that our approach achieves better performance when compared to the state-of-the-art approaches.
\end{abstract}


\section{Introduction}
\label{sec:intro}
Spectrum sensing is a key component of cognitive radio networks (CRNs), essential for enabling dynamic and opportunistic spectrum access~\cite{akyildiz2011,guizani2015large}.
It essentially allows secondary users (\SU s) to know whether and when a licensed band is available prior to using it so as to avoid harming primary users (\PU s).
Due to its vital role, over the last decade or so, a tremendous amount of research has focused on developing techniques and approaches that enable efficient spectrum sensing~\cite{axell2012spectrum,patil2016survey}. Most of the focus has, however, been on single-band spectrum sensing, and the focus on wideband spectrum sensing is more recent and has received lesser attention~\cite{sun2013wideband}.

The key advantage of wideband spectrum sensing over its single-band counterpart is that it allows \SU s to locate spectrum opportunities in wider ranges of frequencies by performing spectrum sensing across multiple bands at the same time. Being able to perform wideband spectrum sensing is becoming a crucial requirement of next-generation CRNs, especially with the emergence of IoT and 5G technologies~\cite{al2016information,niu2015survey,gohil20135g}. This wideband spectrum sensing requirement is becoming even more stringent with FCC's recent new rules for opening up millimeter wave band use for wireless broadband devices in frequencies above 24 GHz~\cite{FCC-mmW-16}.

The challenge, however, with wideband spectrum sensing is that it requires high sampling rates, which can incur significant sensing overhead in terms of energy, computation, and communication.
Motivated by the sparsity nature of spectrum occupancy~\cite{chen2014survey} and in an effort to address the overhead caused by these high sampling rates, researchers have focused on exploiting compressive sampling to make wideband spectrum sensing possible at sub-Nyquist sampling rates~\cite{sharma2016application}.

These research efforts have focused mainly on {\em homogeneous} wideband spectrum, meaning that the entire wideband spectrum is considered as one single block with multiple bands, and the sparsity level is estimated across all bands and considered to be the same for the entire wideband spectrum.
However, in wideband spectrum assignment, applications of similar types (TV, satellite, cellular, etc.) are often assigned bands within the same band block, suggesting that wideband spectrum is {\em heterogeneous}, in the sense that band occupancy patterns are not the same across the different blocks of bands, since different application/user types within each block can exhibit different traffic behaviors. Therefore, sparsity levels may vary significantly from one block to another; this trend has also been confirmed by recent measurement studies~\cite{chen2014survey}.

In this paper, we exploit this inherent, block-like structure of wideband spectrum to design efficient compressive spectrum sensing techniques that are well suited for {\em heterogeneous} wideband spectrum access in {\em noisy} wireless environments. To the best of our knowledge, this is the first work that exploits this spectrum occupancy heterogeneity inherent to wideband spectrum to develop efficient compressive sensing techniques. Specifically, we propose a wideband sensing information recovery algorithm that is more stable and robust than existing approaches. The proposed technique accounts for spectrum occupancy variations across both time and frequency.

We exploit this fine-grained sparsity structure to propose, which to the best of our knowledge, the first spectrum sensing information recovery scheme for {\em heterogeneous} wideband spectrum sensing with {\em noisy} measurements.

\subsection{Our Key Contributions}
In this paper, we make the following contributions:

\begin{itemize}

\item
We propose a weighted $\ell_1-$minimization algorithm that exploits the block-like, sparsity structure of the heterogeneous wideband spectrum to provide an efficient recovery of spectrum occupancy information in noisy CRN environments.
    We design the weights of the algorithm in a way that spectrum blocks that are more likely to be occupied are favored during the search, thereby increasing the recovery performance.

\item We prove that our proposed recovery algorithm outperforms existing approaches in terms of stability and robustness.

\item We derive lower bounds on the probability of spectrum occupation, and use them to determine the sparsity levels that lead to further reduction in the sensing overhead.
\end{itemize}

It is important to mention that our proposed weighted compressive sampling framework, including the derived theoretical results, is not restricted to wideband spectrum sensing applications. It can be applied to any other application where the signal to be recovered possesses block-like sparsity structure. We are hoping that this work
can be found useful for finding efficient solution methodologies to problems (with similar characteristics) in other disciplines and domains.

\subsection{Roadmap}
The remainder of the paper is structured as follows. In Section~\ref{sec:system_model}, we present our system model and the PU bands' occupancy model. Next, our proposed approach along with its performance analysis are presented in Section~\ref{sec:proposed}. The numerical evaluations are then presented in Section~\ref{sec:numerical_results}. Finally, our conclusions are given in Section~\ref{sec:conclusion}.

\section{Wideband Spectrum Sensing Model}
\label{sec:system_model}
In this section, we begin by presenting the studied heterogeneous wideband spectrum model. Then, we present the spectrum sensing preliminaries and setup.

\subsection{Wideband Occupancy Model}
\begin{figure}[t]
\centering{
\includegraphics[width=.8\columnwidth]{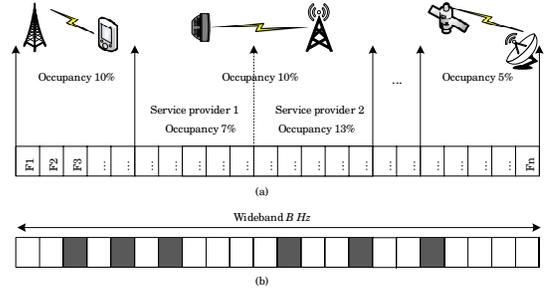}
\caption{$n$ frequency bands occupied by heterogeneous applications with different occupancy rates. The grey bands are occupied by primary users while the white bands are vacant. (a) is the statistical allocation while (b) is a realization of allocation in a given region at a given time slot. }
\label{fig:band_ocup}}
\end{figure}

We consider a heterogeneous wideband spectrum access system containing \n~frequency bands as illustrated by Fig.~\ref{fig:band_ocup}(a).
We assume that wideband spectrum accommodates multiple different types of user applications, where applications of the same type are allocated frequency bands within the same block. Therefore, we consider that wideband spectrum has a block-like occupation structure, where each block (accommodating applications of similar type) has different occupancy behavioral characteristics. The wideband spectrum can then be grouped into $g$ disjoint contiguous blocks, $\mathcal{G}_i, i=1,...,g$,
with $\mathcal{G}_i\bigcap\mathcal{G}_j=\emptyset$ for $i\neq j$. Each block, $\mathcal{G}_i$, is a set of $n_i$ contiguous bands.
Like previous works~\cite{Sun2014TSP},
the state of each band $i$, $\mathcal{H}_i$, is modelled as $\mathcal{H}_i \thicksim \text{Bernoulli}(p_i)$
with parameter $p_i\in[0,1]$ ($p_i$ is the probability that band $i$ is occupied by a \PU). The average number of occupied bands within a block $j$ is then $\bar{k}_j
= \sum_{i\in \mathcal{G}_j } p_i$.

Recall that one of the things that distinguishes this work from others is the fact that we consider a {\em heterogeneous} wideband spectrum; formally, this means that the average number $\bar{k}_j$ of the occupied bands in block $j$ can vary significantly from one block to another. The average occupancies, however, of the different bands within a given block are close to one another; i.e., $p_i\approx p_j$ for all $i,j\in \mathcal{G}_j$. Our proposed framework exploits such a block-like occupancy structure stemming from the wideband spectrum heterogeneity to design efficient compressive wideband spectrum sensing techniques.

\subsection{Secondary System Model}
\begin{figure}
\centering{
\includegraphics[width=.6\columnwidth]{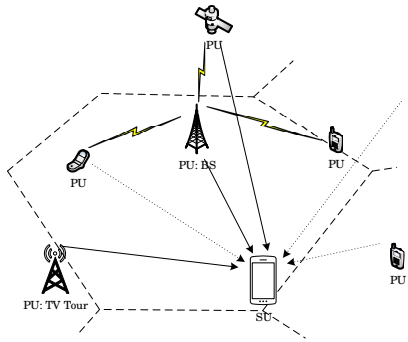}
\caption{A $SU$ performing spectrum sensing over a wideband spectrum. The received signals are coming from the primary users with different levels of energy.}
\label{fig:sys_mod}}
\end{figure}

We consider a SU performing the sensing of the entire wideband spectrum as illustrated by Fig.~\ref{fig:sys_mod}.
The time-domain signal $\boldsymbol{r}(t)$ received by the \SU~can be expressed as
\begin{equation}\label{eqn:signaltime}\nonumber
\boldsymbol{r}(t)=\boldsymbol{h}(t)*\boldsymbol{s}(t)+\boldsymbol{w}(t),
\end{equation}
where $\boldsymbol{h}(t)$ is the channel impulse between the primary transmitters and the SU, $\boldsymbol{s}(t)$ is the PUs' signal, and $\boldsymbol{w}(t)$ is an additive white Gaussian noise with mean $0$ and variance $\sigma^2$.
Ideally, we should take samples with at least twice the maximum frequency, $f^{\max}$, of the signal in order to recover the signal. Let the sensing window be $[0,mT_0]$ with $T_0=1/(2f^{\max})$.
Assuming a normalized number of wideband Nyquist samples per band, then the vector of the taken samples is $\boldsymbol{r}(t)=[r(0),...,r((m_0-1)T_0)]^T$ where $r(i)=r(t)|_{t=iT_0}$ and $m_0=n$.
Note that a reasonable assumption that we make is that the sensing window length is assumed to be sufficiently small when compared to the time it takes a band state to change. That is, each band's occupancy is assumed to remain constant during each sensing time window.

To reveal which bands are occupied, we perform a discrete Fourier transform of the received signal $\boldsymbol{r}(t)$; i.e.,
\begin{equation}\label{eqn:signalfrequency} \nonumber
\boldsymbol{r}_f=\boldsymbol{h}_f\boldsymbol{s}_f+\boldsymbol{w}_f=\boldsymbol{x}+\boldsymbol{w}_f,
\end{equation}
where $\boldsymbol{h}_f$, $\boldsymbol{s}_f$, and $\boldsymbol{w}_f$ are the Fourier transforms of $\boldsymbol{h}(t)$, $\boldsymbol{s}(t)$, and $\boldsymbol{w}(t)$, respectively.
The vector $\boldsymbol{x}$ contains a faded version of the PUs' signals operating in the different bands. Given the occupancy of the bands by their \PU s (as illustrated in Fig.~\ref{fig:band_ocup}(b)) and in the absence of fading and interference, the vector $\boldsymbol{x}$ can be considered {\em sparse}, where sparsity is formally defined as follows.
\begin{definition}
A vector $\boldsymbol{x}\in\mathbb{R}^n$ is said \emph{k-sparse} if it has (or after performing a basis change) at most $k$ non-zero elements~\cite{davenport2011introduction}. That is, $supp(\boldsymbol{x})=\|\boldsymbol{x}\|_{\ell_0} = |\{i : x_i\neq 0\}|\leq k$. The set of $k-$sparse vectors in $\mathbb{R}^n$ are denoted by $\Sigma_{k}=\{\boldsymbol{x}\in\mathbb{R}^n : \|\boldsymbol{x}\|_{\ell_0}\leq k\}$.
\end{definition}
But since, in practice, there will likely be interference coming from other nearby cells and users, the vector $\boldsymbol{x}$ could rather be {\em nearly sparse} than sparse, where nearly sparsity is formally defined next.
\begin{definition}
  A vector $\boldsymbol{x}\in\mathbb{R}^n$ is said \emph{nearly sparse} (called also compressible~\cite{davenport2011introduction}) if most of its components obey a fast power low decay. The $k-$sparsity index of $\boldsymbol{x}$ is then defined as $\sigma_k(\boldsymbol{x},\|.\|_{\ell_p})=\displaystyle{\min_{\boldsymbol{z}\in\Sigma_k}}\|\boldsymbol{x}-\boldsymbol{z}\|_{\ell_p}$.
\end{definition}
Since the wideband spectrum is large, the number of required samples can be huge, making the sensing operation prohibitively costly and the needed hardware capabilities beyond possible. To overcome this issue, researchers have focused on using compressive sampling theory as a way to reduce the number of measurements, given that the wideband spectrum signal to be recovered possesses the sparsity or nearly sparsity property needed to apply such a theory.
After performing the compressive sampling, the resulted signal can be written as
\begin{eqnarray}\nonumber
  \boldsymbol{y} &=& \Psi {\mathcal{F}^{-1}}(\boldsymbol{x} +\boldsymbol{w}_f) \\\nonumber
   &=& {\mathcal{A}}\boldsymbol{x}+\boldsymbol{\eta},
\end{eqnarray}
where $\boldsymbol{y}\in\mathbb{R}^m$ is the measurement vector, ${\mathcal{F}^{-1}}$ is the inverse discrete Fourier transform, and $\Psi$ is the sensing matrix assumed to have a full rank, i.e. $rank(\Psi)=m$. The sensing noise $\boldsymbol{\eta}$ is equal to $\Psi {\mathcal{F}^{-1}}\boldsymbol{w}_f$.

Different from the classical application of compressive sampling for wideband spectrum sensing, in this paper we propose to take advantage of the block-like structure of the occupancy of the wideband spectrum, and design an efficient compressive spectrum sensing algorithm well suited for heterogeneous wideband CRNs. Exploiting the variability of the average band occupancies across the various blocks has the potential for improving the recovery of the wideband spectrum sensing signals, and therefore, the ability of acquiring accurate \PU~detection and spectrum availability information efficiently. In the next section, the proposed wideband spectrum sensing recovery approach will be presented along with its performance analysis.

\section{The Proposed Wideband Spectrum Sensing Information Recovery}
\label{sec:proposed}
The sensing matrix and recovery algorithm are the main challenging components in compressive sampling design. While the former consists of minimizing the number of measurements, the latter consists of ensuring a stable and robust recovery. In this work, we exploit the block-like occupancy structure information of the wideband spectrum to propose a new recovery algorithm that outperforms existing approaches by $1)$ requiring lesser numbers of measurements (better sensing matrix) and $2)$ reducing recovery error (more stable and robust recovery).
In this section, we start by providing some background on signal recovery using classical compressive sampling. Then, we present our proposed approach, and analyze its performance by bounding its achievable mean square errors and its required number of measurements.

\subsection{Background}
In order to acquire spectrum availability/occupancy information, an SU needs first to recover the frequency-domain version of the received signal. Exploiting the fact that the signal is sparse, an ideal recovery can be performed by minimizing the $\ell_0-$norm of the signal. This happens to be NP-hard~\cite{Candestao2005TIT}. It turns out that minimizing the $\ell_1-$norm recovers the sparsest solution with a bounded error that depends on the noise variance and the solution structure~\cite{candes2006stable}. This can be formulated as 
\begin{equation*}
\begin{aligned}
\mathscr{P}_1 : &\; \underset{x}{\text{minimize}}
& & \|\boldsymbol{x}\|_{\ell_1}\\
& \text{subject to}
& & \|\mathcal{A}\boldsymbol{x}-\boldsymbol{y}\|_{\ell_2}\leq \epsilon
\end{aligned}
\end{equation*}
Here, $\epsilon$ is a user-defined parameter chosen such that $\|\boldsymbol{\eta}\|_{\ell_2}\leq \epsilon$. This formulation is known also as Least Absolute Shrinkage and Selection Operator (LASSO)~\cite{candes2006stable}.

Although LASSO is shown to achieve good performance when applied for wideband spectrum sensing recovery, it does not capture, nor exploit the block-like occupancy structure information that is inherent to the heterogeneous wideband spectrum, where the occupancy is homogeneous within each block but heterogeneous across the different blocks of the spectrum.
As we will show later, it is the exploitation of this block-like spectrum occupancy structure that is behind the performance gain achieved by our proposed compressive spectrum sensing recovery algorithm.

\subsection{The Proposed Recovery Algorithm}
Again, in this work, we consider a heterogeneous wideband spectrum that contains $g$ contiguous blocks, $\mathcal{G}_i, i=1,...,g$. Let $\bar{k}_i$ be the average sparsity level of block $\mathcal{G}_i$ (average across all bands belonging to the block), and $\bar{k}$ be the average sparsity level across all blocks.
We assume that the blocks have sufficient different average sparsity levels (otherwise, blocks with similar sparsity levels are merged into one block with a sparsity level corresponding to their average).
These averages are often available via measurement studies, can easily be estimated, or can even be provided by spectrum operators~\cite{mehdawi2015spectrum}.

Intuitively, our key idea consists of incorporating and exploiting the sparsity level variability across the different blocks of the spectrum sensing signal to perform intelligent solution search. We essentially encourage more search of the non-zero elements of the signal $\boldsymbol{x}$ in the blocks that have higher average sparsity levels while discouraging this search in the blocks with low average sparsity levels.
Such variability in the block sparsity levels can be incorporated in the formulation through carefully designed weights. More specifically, we propose the following weighted $\ell_1-$minimization recovery scheme
\begin{equation*}
\begin{aligned}
\mathscr{P}_1^{\omega}: & & \underset{x}{\text{minimize}}
& & \sum_{l=1}^g \omega_l\|\boldsymbol{x}_{l}\|_{\ell_1}\\
& & \text{subject to}
& &  \|\mathcal{A}\boldsymbol{x}-\boldsymbol{y}\|_{\ell_2}\leq \epsilon.
\end{aligned}
\end{equation*}
where $\boldsymbol{x}=[\boldsymbol{x}_1^T,..., \boldsymbol{x}_g^T]^T$, $\boldsymbol{x}_l^T$ is a $n_l\times 1$ vector, and $\omega_l$ is the weight assigned to block $l$ for $l\in\{1,...,g\}$.

The question that arises here is how to design and select these weights.
Generally speaking, given that the average sparsity level differs from one block to another, blocks with higher average sparsity levels are supposed to contain more occupied bands than those blocks with lower averages. This means that if we consider two blocks with two different average sparsity levels, say $\bar{k}_1$ and $\bar{k}_2$, such that $\bar{k}_1<\bar{k}_2$, then to encourage the search for more occupied bands in the second block, the weight $\omega_2$ assigned to the second block should be smaller than the weight $\omega_1$ assigned to the first block. Following this intuition, we set the weights to be inversely proportional to the average sparsity levels. More specifically,
\begin{equation}\label{eqn:weight}
\omega_i=\frac{1/\bar{k}_i}{\sum_{j=1}^g1/\bar{k}_j}~~~~\forall~i \in \{1,...,g\}
\end{equation}

\begin{remark}\emph{Some insights into the proposed scheme}\\
Consider a two-block spectrum with $\bar{k}_1>\bar{k}_2$. For this special case, the recovery algorithm can then be re-written as
\begin{equation*}
\begin{aligned}
\mathscr{P}_1^{\omega,2}: & & \underset{x}{\text{minimize}}
& & \|\boldsymbol{x}\|_{\ell_1}+(\frac{\omega_2}{\omega_1}-1)\|\boldsymbol{x}_{2}\|_{\ell_1}\\
& & \text{subject to}
& &  \|\mathcal{A}\boldsymbol{x}-\boldsymbol{y}\|_{\ell_2}\leq \epsilon.
\end{aligned}
\end{equation*}
Since we are minimizing the $\ell_1-$norm of $\boldsymbol{x}$ and the $\ell_1-$norm of $\boldsymbol{x}_2$, this can be interpreted as ensuring that the vector $\boldsymbol{x}$ is sparse while ensuring that the portion $\boldsymbol{x}_{2}$ of $\boldsymbol{x}$ is also sparse. This means that all solutions that are sparse as a whole but somehow dense in their second portion are eliminated.
\end{remark}

In the remaining of this section, we derive and evaluate the performance achievable by the proposed recovery algorithm by showing that it $1)$ incurs errors smaller than those incurred by existing techniques and $2)$ reduces the sensing overhead by requiring smaller numbers of required measurements.

\subsection{Mean Square Error Analysis}
The following theorem shows that our weighted recovery algorithm incurs, on the average, lesser errors than what $\ell_1-$minimization~\cite{candes2006stable} incurs.

\begin{theorem}\label{theo:error}
Letting $\boldsymbol{x}^{\sharp}$ be the optimal solution for $\mathscr{P}_1^{\omega}$, $\boldsymbol{x}^{\dag}$ the optimal solution for $\mathscr{P}_1$ and $\boldsymbol{y}=\mathcal{A}\boldsymbol{x}_0+\boldsymbol{\eta}$,
we have
\begin{displaymath}
\label{eqn:perf}
E[\|\boldsymbol{x}^{\sharp}-\boldsymbol{x}_0\|_{\ell_2}] \leq E[\|\boldsymbol{x}^{\dag}-\boldsymbol{x}_0\|_{\ell_2}].
\end{displaymath}
\end{theorem}
Note that throughout this paper, we omit the proofs for all the theorems and lemmas for page limitations.
The theorem says that the expected solution to the proposed $\mathscr{P}_1^{\omega}$ is at least as good as the expected solution to $\mathscr{P}_1$. As done by design, it is also expected that the more heterogeneous the wideband spectrum is, the higher the error gap between our proposed algorithm and LASSO is.
This is because the searched solution has the adequate structure captured via the assigned weights.

We now state the following result, which follows directly from Theorem~\ref{eqn:perf}.

\begin{proposition}\label{prop:1}
Our proposed algorithm, $\mathscr{P}_1^{\omega}$, achieves stable and robust recovery\footnote{As defined in~\cite{candes2006stable}, for $\boldsymbol{y}=\mathcal{A}\boldsymbol{x}+\boldsymbol{w}$ such that $\|\boldsymbol{w}\|_{\ell_2}\leq \epsilon$, a recovery algorithm, $\Delta$, and a sensing matrix, $\mathcal{A}$, are said to achieve a stable and robust recovery if there exist $C_0$ and $C_1$ such that
  \begin{equation}\label{eqn:recover}\nonumber
  \|\Delta \boldsymbol{y}-\boldsymbol{x}\|_{\ell_2}\leq C_0 \epsilon+C_1 \frac{\sigma_k(\boldsymbol{x},\|.\|_{\ell_p})}{\sqrt{k}}.
  \end{equation}}.
\end{proposition}

The proposition gives a bound on the incurred error by means of two quantities. The first quantity is an error of the order of the noise variance while the second is of the order of the sparsity index of $\boldsymbol{x}$.

\begin{remark}\emph{Effect of time-variability}\\
We want to iterate that our proposed algorithm is guaranteed to outperform existing approaches on the average, and not on a per-sensing step basis.
This is because although the performance improvement achieved by our technique stems from the fact that blocks with higher average sparsity levels are given lower weights---which is true on the average, it is not unlikely that, at some sensing step, the actual sparsity level of a block with a higher average could be smaller than that of a block with a lower average. When this happens, our algorithm won't be guaranteed to achieve the best performance during that specific sensing step. The good news is that first what matters is the average over longer periods of sensing time, and second,  depending on the gap between the block sparsity averages, this scenario happens with very low probability.

To illustrate, let us assume that the wideband spectrum contains two blocks with average sparsity  $\bar{k}_1=\sum_{j\in \mathcal{G}_1}p_j\approx n_1p_1$ and $\bar{k}_2=\sum_{j\in \mathcal{G}_2}p_j\approx n_2p_2$ with $\bar{k}_2<\bar{k}_1$, where again $|\mathcal{G}_1|=n_1$ and $|\mathcal{G}_2|=n_2$. Here, the occupancy probabilities of all bands in each of these two blocks are assumed to be close to one another. Our approach encourages to find more occupied bands in the first block than in the second block. However, since band occupancy is time varying, then at some given time we may have a lesser number of non-zero components in first block than in the second. This unlikely event, in this scenario, happens with probability
\begin{displaymath}
\sum_{k=1}^{\min(n_1,n_2)}\sum_{l=0}^{k-1}\dbinom{n_1}{l}q_1^l(1-q_1)^{n_1-l}\dbinom{n_2}{k} q_{2}^k(1-q_2)^{n_2-k}
\end{displaymath}
For a sufficiently different average sparsity levels (e.g. having $\bar{k}_1>2\bar{k}_2$), this probability is very low (less than $0.02$).
\end{remark}

Having investigated the design of the recovery algorithm, now we turn our attention to the design of the sensing matrix. The number of measurements, $m$, that need to be taken determines the size of the sensing matrix and hence the sensing overhead of the recovery approach.
Existing approaches determine the required number of measurements by setting the sparsity level to the average number of occupied bands (e.g., $m\geq \bar{k} \log(n/\bar{k})$). However, in wideband spectrum sensing, the number of occupied bands changes over time, and can easily exceed the average number. Every time this happens, it leads to an inaccurate signal recovery (it yields a solution with high error).
To address this issue, in our proposed framework, we do not base the selection of the number of measurements on the average sparsity. Instead, the sparsity level  is chosen in such a way that the likelihood that the number of occupied bands exceeds that number is small. The analysis needed to help us determine such a sparsity level is provided in the next section.

\subsection{PU Traffic Characterization}
Based on the model of occupancy of the wideband provided in the system model, the following lemma gives the probability mass distribution of the number of occupied bands.
\begin{lemma}
The number of occupied bands across the entire wideband has the following probability mass function
\begin{equation}\label{eqn:bingen}\nonumber
  \textrm{Pr}(X=k) = \displaystyle{\sum_{\Lambda\in\mathcal{S}_k}}\Big[ \displaystyle{\prod_{i\in\Lambda}}~p_i\Big]\Big[     \displaystyle{\prod_{j\in\Lambda^c}}(1-p_j)\Big]
\end{equation}
where $\mathcal{S}_k=\{\Lambda:~\Lambda\subseteq \{1,...,n\}, |\Lambda|=k\}$, and $\Lambda^c$ is the complementary set of $\Lambda$.
\end{lemma}

Given this distribution, the average number of occupied bands across the entire wideband spectrum is $\bar{p}=\sum_{i=1}^np_i$. As just mentioned earlier, setting the sparsity level to be fixed to the average $\lfloor\bar{p}\rfloor$ will lead to inaccurate signal recovery, since the likelihood that the number of occupied bands exceeds this sparsity level is not negligible.
In the following theorem, we provide a lower bound on the probability that the number of occupied bands is below an arbitrary sparsity level.

\begin{theorem}\label{theo:bound}
  The probability that the number of occupied bands is below a sparsity level $k_0$ is low bounded by
  \begin{eqnarray}\label{eqn:bound}\nonumber
    \textrm{Pr}(X\leq k_0) &=&\sum_{k=0}^{k_0}\displaystyle{\sum_{\Lambda\in\mathcal{S}_k}}\Big[     \displaystyle{\prod_{i\in\Lambda}}~p_i\Big]\Big[     \displaystyle{\prod_{j\in\Lambda^c}}(1-p_j)\Big]\\
    &\geq& 1-\frac{e^{k_0-\sum_{i}^np_i}}{(k_0/\sum_{i}^np_i)^{k_0}}
  \end{eqnarray}
\end{theorem}
Since the sparsity level is a time-varying process, this theorem gives a probabilistic bound on how to choose a sparsity level such that the level will be exceeded only with a certain probability. Now depending on the allowed fraction, $\alpha$, of instances in which the actual number of occupied bands exceeds the sparsity level, Theorem~\ref{theo:bound} can be used to determine the sparsity level, $k_0$, that can be used to determine the required number of measurements, $m$, such that $m=\mathcal{O}(k_0\log(n/k_0))$. For example, if $\alpha$ is set to $5\%$, then it means that only about $5\%$ of the time the actual number of occupied bands exceeds the number $m$. As expected, there is a clear tradeoff between $\alpha$ and $m$. Smaller values of $\alpha$ requires higher values of $m$, and vice-versa. In our numerical evaluations given in the next section, $\alpha$ is set to $4\%$.

\section{Numerical Evaluation}
\label{sec:numerical_results}
In this section, we evaluate our proposed wideband spectrum sensing approach and we compare its performance to the state-of-the-art approaches.
Consider a primary system operating over a wideband consisting of $n=256$ bands. We assume that the wideband contains $g=4$ blocks with equal sizes. The average probabilities of occupancy in each block are as follows: $\bar{k}_1=0.1\times 64$, $\bar{k}_2=0.01\times 64$, $\bar{k}_3=0.1\times 64$, $\bar{k}_4=0.01\times 64$.
To model the signals coming from the active users, we generate them in the frequency domain with random magnitudes (which captures the effect of the different channel SNRs that every operating PU has with the SU).
At the SU side, the sensing matrix $\Psi$ is generated according to a Bernoulli distribution with zero mean and $1/m$ variance. We opted for a sub-Gaussian distribution since it guarantees the RIP with high probability~\cite{davenport2011introduction}. Here, the number of measurements is generated first according to $m=\mathcal{O}(k_0\log(n/k_0))$.

\begin{figure}
\centering{
\includegraphics[width=.8\columnwidth]{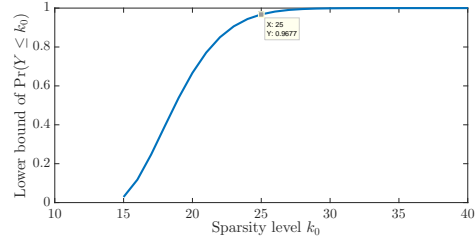}
\caption{The Lower bound of $\textrm{Pr}(X<k_0)$ as a function of the sparsity level $k_0$. }
\label{fig:sparsity}}
\end{figure}

We fix $k_0$ to $25$ which according to Theorem~\ref{theo:bound} is satisfied with a probability that exceeds $0.96$ (Fig. \ref{fig:sparsity}). Now assuming an RIP constant $\delta_{2k_i}\leq1/2$ and replacing $k_0$ and the RIP constant with their values in Theorem 3 yields that the number of measurements should be at least $29$.
We use CVX for the solving of the optimization problem~\cite{grant2008cvx}.

\begin{figure}
\centering{
\includegraphics[width=.9\columnwidth]{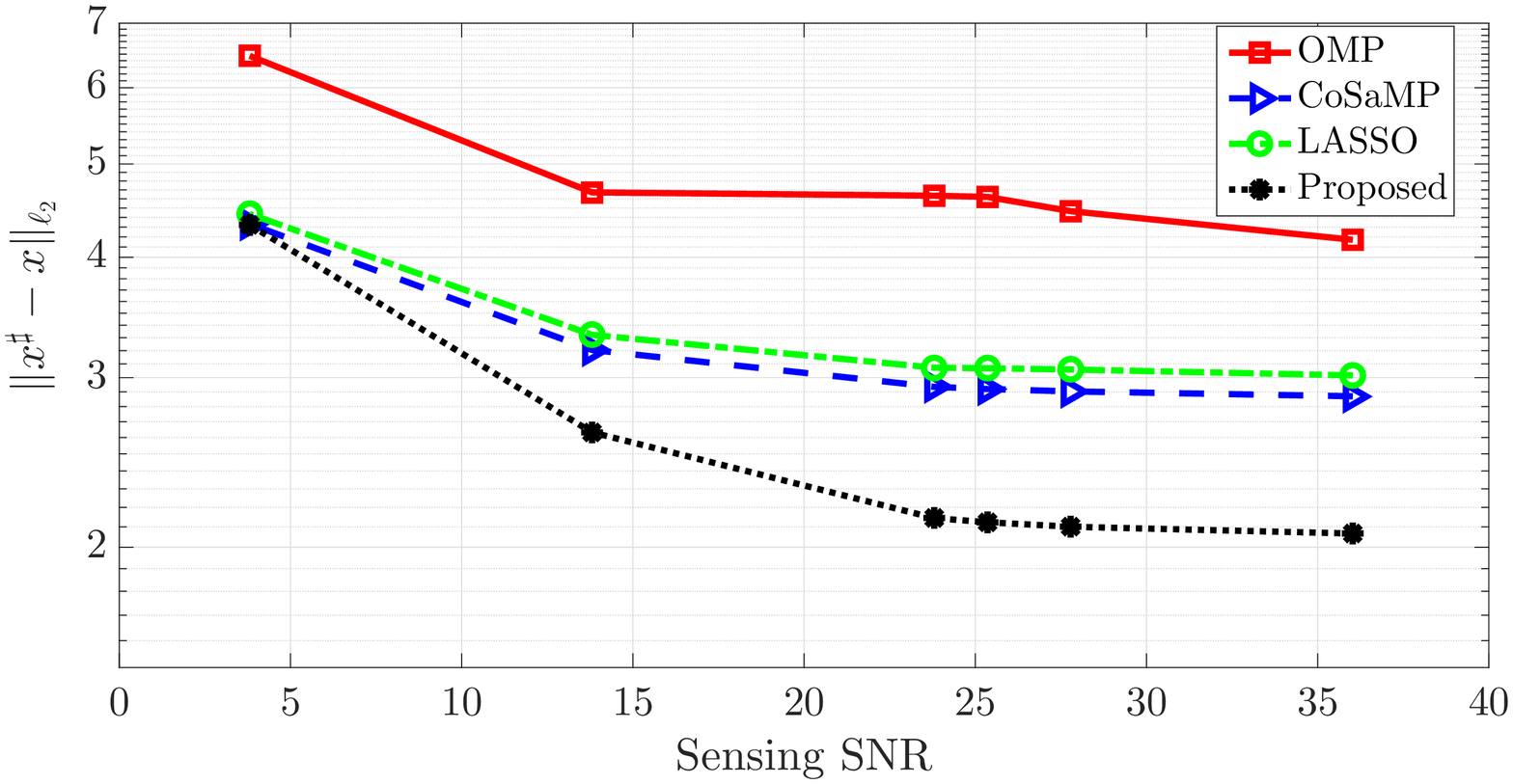}
\caption{Comparison between the recovery approaches in terms of mean square error as a function of the sensing SNR ($m=27$).}
\label{fig:perf1}}
\end{figure}

A first performance that we look at is the mean square error $\|\boldsymbol{x}^{\sharp}-\boldsymbol{x}_0\|_{\ell_2}$ as a function of the sensing SNR defined as $\textrm{SNR}=\frac{\|\mathcal{A}\boldsymbol{x}\|_{\ell_2}^2}{\|\boldsymbol{\eta}\|_{\ell_2}^2}$,
where $\|\mathcal{A}\boldsymbol{x}\|_{\ell_2}^2=(\mathcal{A}\boldsymbol{x})^T\mathcal{A}\boldsymbol{x}$ and $\|\boldsymbol{\eta}\|_{\ell_2}^2=\boldsymbol{\eta}^T\boldsymbol{\eta}$.
In Fig.~\ref{fig:perf1}, we compare our proposed technique to the existing approaches. Compared to LASSO~\cite{candes2006stable}, CoSaMP~\cite{needell2009cosamp}, and (OMP)~\cite{tropp2007signal}, our proposed approach achieves a lesser error when fixing the number of measurement $m$ to $35$. This is because we account for the average sparsity levels in each block, thereby favoring the search on the first and third block rather than the two others. Also, observe that as the sensing SNR gets better, not only does the error of the proposed technique decrease, but also the error gap between our technique and that of the other ones increases. This is because the noise effect becomes limited. Furthermore, OMP has the worst performance as it requires higher number of measurements to perform well.

\begin{figure}
\centering{
\includegraphics[width=.9\columnwidth]{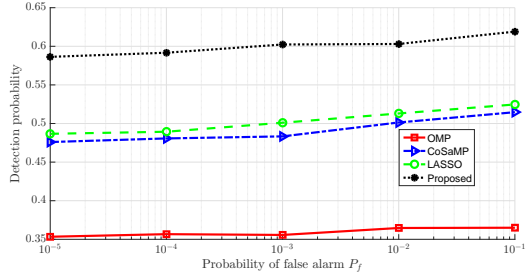}
\caption{Probability of detection as a function of the probability of false alarm with a reduced number of measurements ($m=27$ and sensing SNR$=16.5~dB$).}
\label{fig:detec}}
\end{figure}

After recovering the signal and in order to decide on the availability of the different bands, we compare the energy of the recovered signal in every band with a threshold~\cite{digham2007energy}, $ \lambda=\frac{\mathbb{E}(\|\boldsymbol{\eta}\|_{\ell_2}^2)}{m}\Big(1+\frac{Q^{-1}(P_f)}{\sqrt{n/2}}\Big)$
where $P_f$ is a user-defined threshold for the false alarm probability. It is defined as the probability that a vacant band is detected as occupied, and is expressed as $\frac{1}{\sum_{i=1}^n\mathcal(1-{H}_i)}\sum_{i=1}^nPr(|x_i|^2
\geq \lambda|\mathcal{H}_i=0)$.
${Q^{-1}}$ is the inverse of the $Q-$function. In Fig.~\ref{fig:detec}, we plot this detection probability as a function of the false probability for a fixed average sensing SNR, where the detection probability is computed as $\frac{1}{\sum_{i=1}^n\mathcal{H}_i}\sum_{i=1}^nPr(|x_i|^2\geq \lambda |\mathcal{H}_i=1)$.

\section{Conclusion}
\label{sec:conclusion}
In this work, we proposed an efficient wideband spectrum sensing technique based on compressive sampling. We proposed a weight recovery approach that accounts for the block-like structure inherent to the heterogeneous  nature of wideband spectrum allocation. We showed that the proposed approach outperforms existing approaches by achieving lower mean square errors and enabling higher detection probability when compared to the-state-of-the-art approaches.

\section{Acknowledgment}
This work was made possible by NPRP grant \# NPRP~$5-319-2-121$ from the Qatar National Research Fund (a member of Qatar Foundation). The statements made herein are solely the responsibility of the authors.
\bibliographystyle{IEEEtran}
\bibliography{References}
\end{document}